\documentclass[fleqn,usenatbib]{mnras}

\usepackage{newtxtext,newtxmath}

\usepackage[T1]{fontenc}

\DeclareRobustCommand{\VAN}[3]{#2}
\let\VANthebibliography\thebibliography
\def\thebibliography{\DeclareRobustCommand{\VAN}[3]{##3}\VANthebibliography}

\usepackage{graphicx}	
\usepackage{amsmath}	

\title[LHAASO protons versus LHAASO diffuse gamma-rays: a consistency check]{LHAASO protons vs LHAASO diffuse gamma-rays}

\author[L. E. Espinosa Castro et al.]{L. E.~Espinosa Castro,$^{1,2}$\thanks{luis.espinosacastro@gssi.it}
F. L.~Villante$^{2,3}$, 
V. Vecchiotti$^{4}$, 
C. Evoli,$^{1,2}$
G. Pagliaroli$^{2}$\thanks{giulia.pagliaroli@lngs.infn.it}
\\
$^{1}$Gran Sasso Science Institute, Viale Francesco Crispi 7, 67100, L’Aquila, Italy\\
$^{2}$INFN, Laboratori Nazionali del Gran Sasso, Via G. Acitelli 22, 67100, Assergi, Italy\\
$^{3}$Department of Physical and Chemical Sciences,
University of L’Aquila, 67100 L’Aquila, Italy\\
$^{4}$Department of Physics, NTNU, NO-7491 Trondheim, Norway
}

\date{Accepted 2025 July 28. Received 2025 July 15; in original form 2025 June 17}

\pubyear{\the\year{}}

\begin{document}
\label{firstpage}
\pagerange{L20--L26}
\maketitle

\begin{abstract}
We perform the first direct consistency check between the recently measured proton spectrum at the knee by Large High
Altitude Air Shower Observatory (LHAASO) and the collaboration’s own high-precision mapping of Galactic diffuse gamma-ray emission. By modeling the hadronic gamma-ray production using the updated cosmic-ray spectra, gas templates and cross-section models, we show that the predicted gamma-ray flux robustly overshoots the LHAASO data in both inner and lateral Galactic regions. This persistent mismatch in both normalization and spectral shape challenges conventional scenarios linking the local cosmic-ray sea to Galactic gamma-ray emission, and calls for a revision of current cosmic ray models in the TeV-PeV sky.
\end{abstract}



\section{Introduction}
\label{sec:intro}

Recently, the Large High Altitude Air Shower Observatory (LHAASO) measured the cosmic-ray (CR) proton spectrum with unprecedented accuracy in the energy range known as the ‘knee'~\citep{Cao_2025b}. 
A few months ago, the same collaboration also provided an updated measurement of high-energy (TeV-PeV) Galactic diffuse gamma-ray emission~\citep{Cao_2023,Cao_2025a} using in combination the Water Cherenkov Detector Array (WCDA) and the Square Kilometer Array (KM2A). This diffuse gamma-ray emission is primarily generated by the interaction of cosmic rays (mainly protons and helium) with the gas present in our Galaxy.

The location of the knee, around a few PeV, has long been associated with the maximum energy achievable by Galactic CR accelerators, and its precise shape carries crucial information about source spectra, propagation effects, and possible contributions from different nuclear species. LHAASO’s new proton spectrum, with its reduced statistical and systematic uncertainties, offers an opportunity to revisit existing models of CR propagation at the end of the Galactic spectrum. At the same time, the diffuse gamma-ray map measured by LHAASO traces the product of the local CR intensity and the interstellar gas distribution, providing a complementary probe of the same underlying particle population. In particular, any mismatch between the inferred CR proton flux and the secondary gamma-ray emission would point toward revisions in our understanding of, e.g., diffusion coefficients \citep{Ptuskin_1993, Syrovatskii_1971}, different gas densities along the Galactic Plane, contributions from unresolved source populations~\citep{Steppa2020aa,Cataldo:2020qla,Vecchiotti:2021yzk,Lipari:2024pzo,He2025apj}, etc.

Several theoretical studies have modeled the TeV-PeV diffuse gamma-ray flux by assuming a smooth CR sea normalized to locally measured spectra~\citep{Lipari_2018,Cataldo:2019qnz,Schwefer:2022zly,Vecchiotti:2024kkz}, or by exploring how CR transport models, incorporating interactions with interstellar turbulence, affect this observable~\citep{Lipari_2018,Cataldo:2019qnz,Gaggero_2015,DeLaTorre_2025}. 
Until now, however, such comparisons have been limited by either the accuracy of the proton measurements or the angular and energy resolution of the gamma-ray observations. LHAASO’s simultaneous, high-precision observations remove many of these previous ambiguities. By convolving the measured proton spectrum with a standard $\pi^0$-decay emissivity model and folding in the latest gas survey data, one can generate a ``predicted'' diffuse gamma-ray intensity that is directly comparable to the LHAASO gamma-ray dataset.

Accordingly, evaluating the consistency between these two measurements is of critical importance. In this work, we analyze the implications of the newly released CR proton spectrum on the predicted secondary diffuse gamma-ray flux and determine the conditions under which both datasets can be brought into agreement.

\section{The cosmic-ray spectrum at the 'knee'}
\label{sec:CR}

Motivated by the recent discovery of rigidity‐dependent spectral breaks in nearly all CR species by direct missions (PAMELA~\citealt{Adriani_2011}, AMS-02~\citealt{Aguilar_2015a, Aguilar_2015b}, DAMPE~\citealt{An_2019, Alemanno_2021}, CALET~ \citealt{Adriani_2019, Adriani_2023}, CREAM~\citealt{Yoon_2017} and ATIC-2~\citealt{Panov_2009}) we adopt the parameterization introduced in~\citet{Antonyan_2000} to describe the proton and helium fluxes as:
\begin{equation}
    \label{eq:cr_spectrum}
    \phi_{A}(E) = \left[ K_A
      \left(\frac{E}{E_0}\right)^{-\alpha_{1}(A)} \right] S_{A}(E) \, , 
\end{equation}
where the label $A$ (here $A = $ H or He) refers to the considered nucleus, and $K_A$ is a normalization factor with units (GeV cm$^2$ s sr)$^{-1}$ at an arbitrary energy $E_0$. 
On the other hand, the function $S_A(E)$, given by 
\begin{equation}
    \label{eq:cr_shape}
    S_{A}(E) = \Pi_{i}
    \left[1+\left(\frac{E}{E_{b,i}(A)}\right)^\frac{1}{\omega_i(A)}\right]^{-\Delta \alpha_{i}(A) \, \omega_i(A)}
\end{equation}
models the spectrum at larger energies by introducing a series of breaks located at energies $E_{b,i}(A)$ with spectral index changing by $\Delta \alpha_i(A) = \alpha_{i+1} (A) - \alpha_{i}(A)$ over an energy-width $\omega_i(A)$ in logarithmic scale.

\par As it is discussed, e.g., in \citet{Lipari_2020}, CR proton data by the aforementioned direct detection experiments reveal two spectral breaks at $\sim 670$ GeV and $\sim 16$ TeV. To properly account for diffuse $\gamma-$ray emission measured by LHAASO, however, it is necessary to extend the analysis to higher proton energies. Indeed, the most relevant energy range for CR protons contributing to $\gamma-$ray diffuse emission measured by LHAASO spans from $10^4$ to $10^7$ GeV. This follows from the fact that a photon with energy $E_\gamma$ is most probably produced by a nucleon with energy $E_{\rm n} \sim 10\, E_\gamma$ in CR hadronic interactions~\citep{Kelner:2006tc}. 
\par
\begin{figure}
\centering
\includegraphics[scale=0.35]{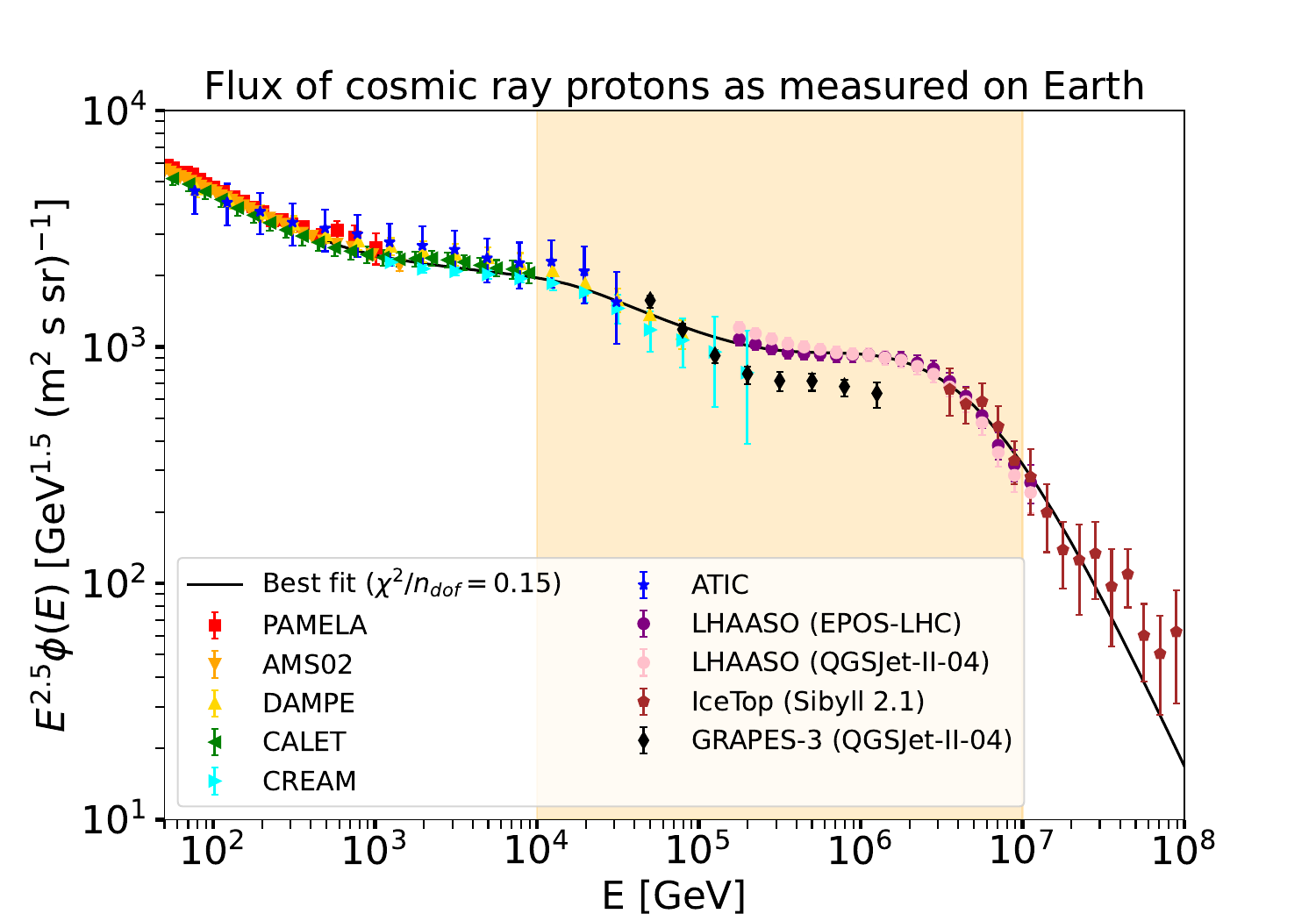}
\centering
\caption{Flux of Galactic cosmic ray protons as a function of energy per nuclei. Observation data by direct detection experiments (PAMELA \citet{Adriani_2011}, AMS-02 \citet{Aguilar_2015a, Aguilar_2015b}, DAMPE \citet{An_2019, Alemanno_2021}, CALET \citet{Adriani_2019, Adriani_2023}, CREAM \citet{Yoon_2017} and ATIC-2 \citet{Panov_2009}) as well as ground-based observatories (GRAPES-3 \citet{Varsi_2024}, IceTop \citet{Aartsen_2019}, and LHAASO \citet{Cao_2025b}) shown with colored scatter points. Best-fit of our model represented with solid black line. Shaded area correspond to the energy range of interest for the production of TeV-PeV $\gamma$-rays.}
\label{fig:cosmic_ray_fluxes}
\end{figure}

 Within this energy range, highlighted by shaded bands in Fig.~\ref{fig:cosmic_ray_fluxes}, CR proton data show the existence of two additional spectral breaks. The first one at $\sim10^2$ TeV was firstly observed by the indirect experiment GRAPES-3 \citep{Varsi_2024} and it is now more precisely constrained by the results reported by the LHAASO collaboration. The second corresponds to the cosmic ray proton knee, located at around $3$ PeV, finally probed by LHAASO. 

Considering the above discussion, we fit CR proton data below $10^8$~GeV by using a functional form featuring four spectral breaks. The resulting fit is shown with a black solid line in Fig.~\ref{fig:cosmic_ray_fluxes} while the best fit parameter values are reported in the Supplementary Material. Our fit includes the LHAASO measurements in the energy range from $0.158$ to $12.6$ PeV (we consider the measurements obtained with the EPOS-LHC hadronic model). Notably, these observations are in good agreement with the previous IceTop data~\citep{Aartsen_2019} above the knee, and statistically dominate the determination of the proton spectrum at PeV energies.
\par Furthermore, the LHAASO data extend to lower energies than those covered by IceTop, enabling, for the first time, to overlap CR proton measurements from GeV energies up to the \textit{knee} region.  
We remark a disagreement between the new LHAASO datasets and the observations reported by the GRAPES-3 experiment at lower energies. Therefore, we discard the latter for the determination of the best-fit proton flux.



\par The CR helium flux is determined by following a similar procedure, i.e., we fit the available data for this component by using the functional form given in Eq.~\ref{eq:cr_spectrum}. Our results are shown in the Supplementary Material, alongside the best-fit parameters. The CR helium flux is comparable to the proton flux at low energies. However, it becomes dominant above $10^7$~GeV, due to the fact that the helium knee is located at a larger energy than the proton knee. 

\par Interestingly, the spectral breaks in the helium flux are found, within uncertainties, at roughly twice the energy of the corresponding breaks in the proton spectrum. This alignment suggests that the spectral features of both elements appear at nearly the same rigidity $R$, as expected if the origin of the breaks is governed by rigidity-dependent processes such as acceleration or transport.
 

%
Based on this observation, we model the spectra of heavier elements by assuming they share the same rigidity dependence as helium, differing only by a normalization factor. This implies that the energy spectrum of a heavier nucleus is
$\phi_{A}(E) \propto \, \phi_{\rm He}( \xi_A \, E ) $, i.e., it is shifted to larger energies with respect to that of helium by a multiplicative factor that is determined by the nuclear charge $Z_A$ or mass number $A$ according to $\xi_A \equiv Z_{\rm He}/Z_{A} \simeq A_{\rm He}/A$. In this assumption, in the low-energy regime where the spectrum
can be approximated by a power law with index $\alpha_{1, \text{He}}$, the total heavy element contribution (relative to that of helium) is roughly constant and given by $ \phi_{\rm heavy}(E)/ \phi_{\rm He}(E)=  \sum_{A>4} (K_{A} / K_{\rm He})$.

\par This contribution can be estimated from observations of the
all particle flux $\phi_{\rm  tot}(E)$ measured by LHAASO \citep{Cao_2024} and
other detectors (HAWC \citet{Morales-Soto_2021}, Tunka \citet{Prosin_2014}, KASCADE \citet{Antoni_2005, Finger_2011} and IceTop \citet{Aartsen_2019}). In particular, we can write $\phi_{\rm heavy}(E) = \phi_{\rm tot}(E) - \phi_{\rm p}(E) - \phi_{\rm  He}(E)$, allowing us to conclude that $\phi_{\rm heavy}(E)/\phi_{\rm He}(E) =\eta\sim 1.7$ at low energies directly
from observational data. Finally, we note that the assumption of a common rigidity spectrum naturally
predicts that heavier elements become increasingly significant at higher energies 
since the location of their spectral knees is shifted in energy proportionally to their mass number.
\par Although helium and heavy elements play a relevant role in shaping the CR all-particle spectrum, they are much less significant for assessing the diffuse $\gamma-$ray flux. Indeed, $\gamma-$ray production by hadronic interactions is primarily
determined at these energies by the total CR nucleon flux, which is given by
\begin{equation}
\label{eq:cr_nucleon}
 \phi_{\rm CR}(E_{n}) = \sum_{A} A^2 \, \phi_{A}(A E_{n}) 
\end{equation}
where $E_{n}$ is the energy per nucleon. The factor $A^2$ comes from considering that the number of nucleons ($N_n$) in a nucleus of a atomic mass $A$ is $N_n=A \, N_i$, while the energy per nucleon is $E_n=E/A$. Therefore, the differential flux of nucleons scales as $\phi(E_n)\sim dN_n/dE_n \propto A^2 dN_A/dE$.
\par Since the fluxes $\phi_A(E)$ of the various nuclear species decrease with their total energy faster than $E^{-2}$, the combined effect of the prefactor $A^2$ and of the shift in energy $E=A E_{\rm n}$ is to suppress helium and heavy element contributions (at a given nucleon energy)  with respect to that of hydrogen. Moreover, helium and heavy elements, having the same rigidity spectrum and a common mass-to-charge ratio ($A/Z_A\simeq 2$),
have breaks in their spectra located  at identical nucleon energies and all of them shifted to lower energy by approximately a factor 2 with respect to those of protons (for which $A_H/Z_H\simeq 1$).
\par Taking all this into account, the CR nucleon flux can be written as:
\begin{equation}
\label{eq:cr_all_nucleon}
 \phi_{\rm CR,\odot} (E_{\rm n}) =  \phi_{\rm p}(E_{\rm n}) + (1+k) \, \phi_{\rm He,
   \, CR}(E_{\rm n})
\end{equation}
where $\phi_{\rm He, \rm CR}(E_{\rm n}) = A_{\rm He}^2 \, \phi_{\rm He}(A_{\rm He} E_{\rm n})$ is the helium contribution to nucleon flux and the factor
\begin{equation}
\label{eq:heavy_factor}
k = \sum_{A>4} \left(\frac{A}{A_{\rm He}}\right)^{2-\alpha_1(He)} \frac{K_{A}}{K_{\rm He}}
\end{equation}
takes into account heavy elements contributions. The determination of $K_A$ requires the knowledge of the CR composition. However, we can obtain an estimate for this parameter by assuming that one element dominates the all-particle spectrum. In this assumption, the determination of $\eta=\frac{K_{A}}{K_{\rm He}} \sim 1.7$ at low energies translates into:
\begin{equation}
\label{eq:heavy_factor_limit}
k = \eta \left(\frac{A}{A_{\rm He}}\right)^{2-\alpha_1(He)}
\end{equation}
from which we see that lighter elements are expected to give a larger contribution to CR nucleon spectrum. We obtain a conservative range for $k$ by taking $A=12$ (carbon) and $A=56$ (iron) as extreme values.
\par
\begin{figure}
\centering
\includegraphics[scale=0.35]{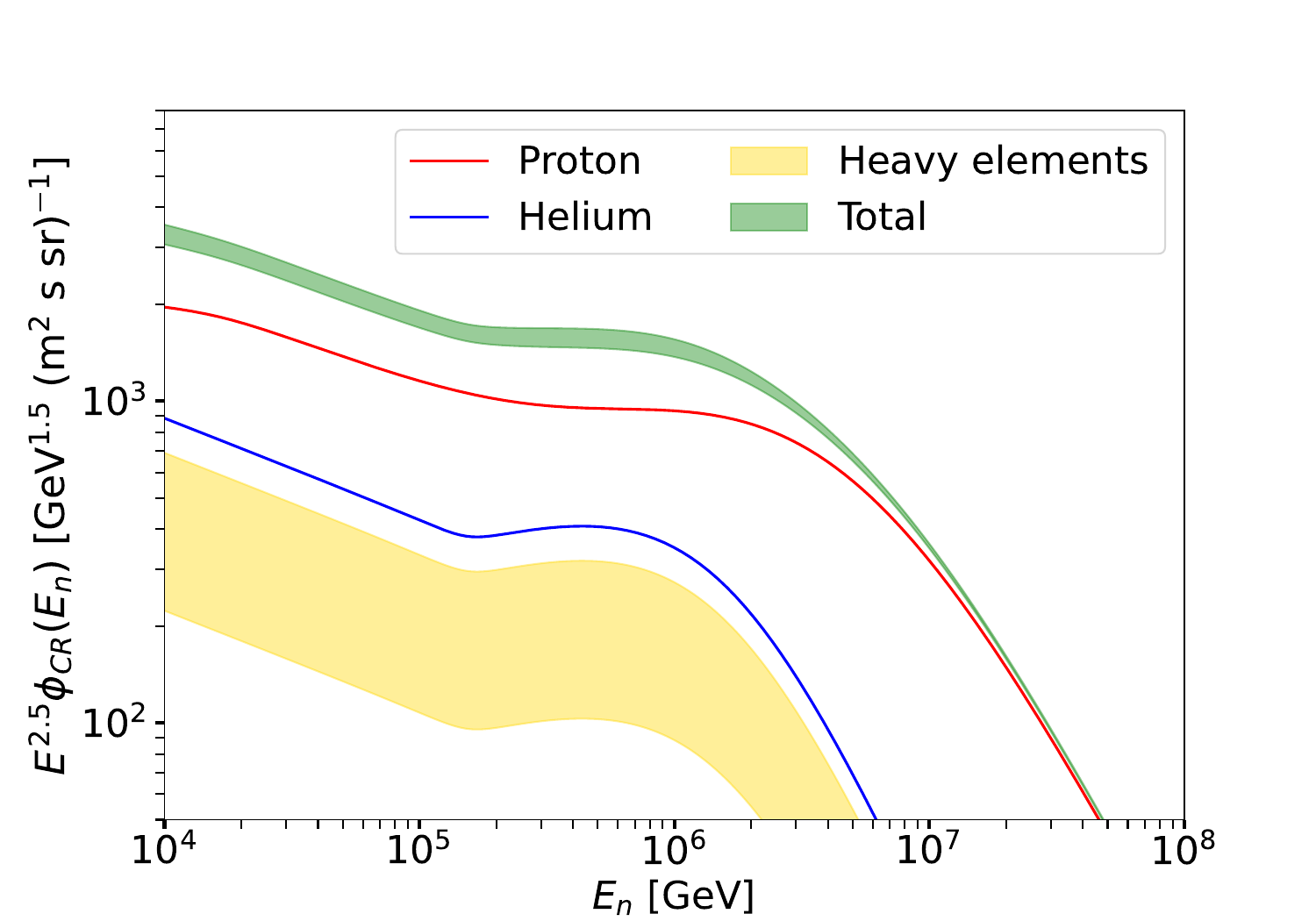}
\caption{Flux of Galactic cosmic ray protons (red line) and helium (blue line) as a function of energy per nucleon. Heavy component added with yellow shaded band within the minimum (iron domination) and maximum (carbon domination) limits. Total flux CR nucleon flux represented with green shaded band.}
\label{fig:cosmic_ray_components}
\end{figure}

 The different contributions to the CR nucleon flux are reported in Fig \ref{fig:cosmic_ray_components}, where the heavy element contribution is given as a shaded band delimited by the maximum and minimum limits described above. Contrary to what happens for the all-particle CR spectrum, helium and heavy elements are subdominant everywhere and, in particular, have a negligible role at PeV
energies. 
\par All this shows that $\gamma-$ray diffuse emission is essentially probing the CR proton spectrum and in particular the position of the CR proton knee. For this reason, the new determination of CR proton flux provided by LHAASO, which play a dominant role in constraining the proton spectrum up PeV energies, has a key role in this analysis and radically improve the robustness of the conclusion that can be obtained. 

\section{The Galactic diffuse gamma-ray emission}
\label{sec:gamma}

The diffuse $\gamma$-ray emission resulting from CR interactions with the interstellar medium can be expressed as in \citet{Pagliaroli:2016, Cataldo:2019qnz}:
\begin{eqnarray}
\label{eq:gamma_flux}
\phi_{\gamma} (E_\gamma,\hat{n}_\gamma) &=&
\int_{E_\gamma}^{\infty} d E_{\rm n} \,
\frac{d\sigma}{dE_\gamma}\left(E_{\rm n},\, E_\gamma \right)\,\\
& & \nonumber
\hspace{-2.5cm}
\int_{0}^{\infty}  dl \,\phi_{\rm CR} (E_{\rm n},{\bf r}_{\odot}+ l \, \hat{n}_\gamma )\,
n_{g} ({\bf r}_{\odot}+l \, \hat{n}_\gamma )\, e^{-\tau(l, E_\gamma)}.
\label{Eq:DiffFlux}
\end{eqnarray}
Here, $d\sigma/dE_\gamma$ is the differential cross section for $\gamma$-ray production in nucleon-nucleon collisions, $n_{g}({\bf r})$ is the gas density distribution, $ r_{\odot}=8.5$~kpc is the position of the Sun and $\tau(l,E_\gamma)$ represents the optical depth due to pair production on background radiation field photons.

As evident from this equation, calculating the $\gamma$-ray flux 
 requires the knowledge of the CR nucleon flux $\phi_{\rm  CR} (E_{\rm
   n},{\bf r})$ in all the region of the Galaxy where the gas density
 is significant. We define the ratio:
 \begin{equation}
 \mathcal R (E_{\rm
   n},{\bf r}) \equiv
  \frac{ \phi_{\rm  CR} (E_{\rm
   n},{\bf r})}{\phi_{\rm  CR,\odot} (E_{\rm
   n})}
\end{equation} 
which measures the CR flux at a generic point relative to that at the
Sun's position.
Generically, this ratio may vary with both energy and position.
However, standard propagation models (e.g., GALPROP) predict that the CR spectrum does not depend on the position in the Galaxy apart from a normalization factor, i.e., $\mathcal R (E_{\rm n},{\bf r}) = g({\bf r})$.
In this assumption, the gamma-ray flux can be expressed as:
\begin{equation}
\phi_{\gamma} (E_\gamma,\hat{n}_\gamma) = {\mathcal
  N}(E_\gamma,\hat{n}_\gamma) \,
{\mathcal
  Y}(E_\gamma)
\end{equation}
where
\begin{equation}
{\mathcal
  Y}(E_\gamma) = \int_{E_\gamma}^{\infty} d E_{\rm n} \,
\frac{d\sigma}{dE_\gamma}\left(E_{\rm n},\, E_\gamma \right)\, \phi_{\rm CR,\odot} (E_{\rm n})
\end{equation}
represents the gamma-ray yield produced by the local CR flux discussed in the previous section, and
\begin{equation}
{\mathcal N}(E_\gamma,\hat{n}_\gamma)  =   \int_{0}^{\infty}  dl\,
g ({\bf r}_{\odot}+l \, \hat{n}_\gamma )
n_{g} ({\bf r}_{\odot}+l \, \hat{n}_\gamma )\, e^{-\tau(l, E_\gamma)}
\end{equation}
encodes the information about gas and CR distribution across the Galaxy.

\par The function ${\mathcal Y}(E_\gamma)$ depends solely on the photon energy and is computed using the AAFRAG parameterization~\citep{Koldobskiy_2021, Kachelriess:2022khq} for the photon production cross section, which we find to be in closest agreement with the approach of~\citet{Orusa:2023azt}, based entirely on accelerator data fits. A comparison with alternative cross-section models is provided in the discussion section.

The function ${\mathcal  N}(E_\gamma,\hat{n}_\gamma)$ may depend both on the photon energy and arrival direction. 
The energy dependence is typically weak and it is due to the fact that the optical
depth $\tau(l,\, E_{\gamma})$ depends on the photon energy. 
Since the main target is provided by CMB radiation with typical
energies at the level of $\sim 10^{-4}$~eV,  the kinematical threshold for pair production implies that only photons with energy larger than $\sim {\rm PeV}$ can be affected.

Directional dependence, by contrast, is mainly determined by the distribution of gas and CRs in the Galaxy.
The interstellar gas is mainly composed of atomic (H) and molecular
hydrogen (H$_2$), whose distributions are traced by the 21-cm
\citep{hi4pi2016} and CO \citep{Dame2001} emission lines.
We include both components in our analysis by the map provided by the GALPROP code\footnote{\href{http://galprop.stanford.edu/}{galprop.stanford.edu}} \citep{Porter_2022}. 
Alternatively, the gas distribution can be inferred from the dust opacity
($\tau_{\rm D}$, obtained at 353 GHz), measured by Planck
Collaboration\footnote{\href{https://www.esa.int/Science_Exploration/Space_Science/Planck}{esa.int/Science$\_$Exploration/Space$\_$Science/Planck}}
\citep{Planck:2016frx}.
This quantity can be used as a tracer for the hydrogen gas column
density ($N_{\rm H}$) since dust is uniformly mixed with neutral gas.
The dust-to-gas conversion factor is calibrated on experimental data; for our analysis, we use \(X^{-1}_{\rm D} \equiv \left(\frac{\tau_{\rm D}}{N_{\rm H}}\right) = 1.18 \times 10^{-26} \, \text{cm}^2\), as reported by \citet{Planck2011}. 
To account for heavier elements, we scale the
hydrogen density by a factor of 1.42 in both models, reflecting the
Solar System composition, assumed to be representative of the
entire Galactic Disk \citep{Ferriere:2001rg}.

Finally, the spatial profile $g({\bf r})$ reflects the distribution
of CRs in the Galaxy.
Its specific form is determined by the spatial distribution of high
energy CR sources  and by CR propagation in the Galactic magnetic
field, see e.g., \citet{Cataldo:2019qnz} for approximate expressions.
For simplicity, in this work we initially assume that CR are 
uniformly distributed in the Galaxy (i.e. $g({\bf r}) =1$) and later address how potential
deviations from this assumption could impact our results.
Under this assumption, the function ${\mathcal
  N}(E_\gamma,\hat{n}_\gamma)$ essentially describes the gas column
density along a given line of sight.
It is important to note that any deviation from $g({\bf r})=1$ would
alter only the angular structure of the diffuse gamma-ray flux, not
its energy spectrum.
Therefore, spectral discrepancies between the predicted and observed gamma-ray fluxes can serve as indicators of a position-dependent CR spectrum in the Galaxy.

\section{Results}
\label{sec:res}
\begin{figure}
\centering
\includegraphics[scale=0.33]{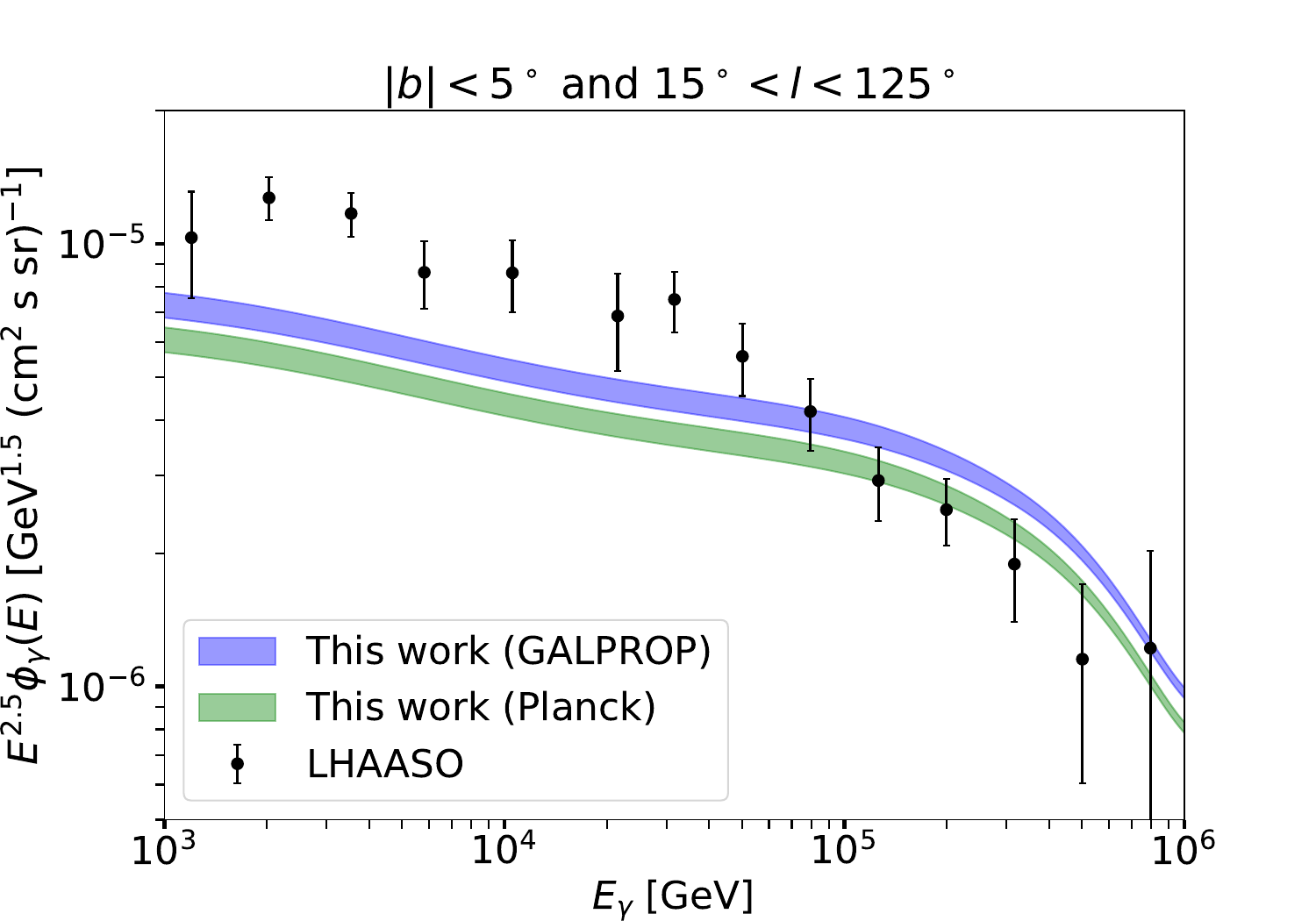}
\centering
\includegraphics[scale=0.33]{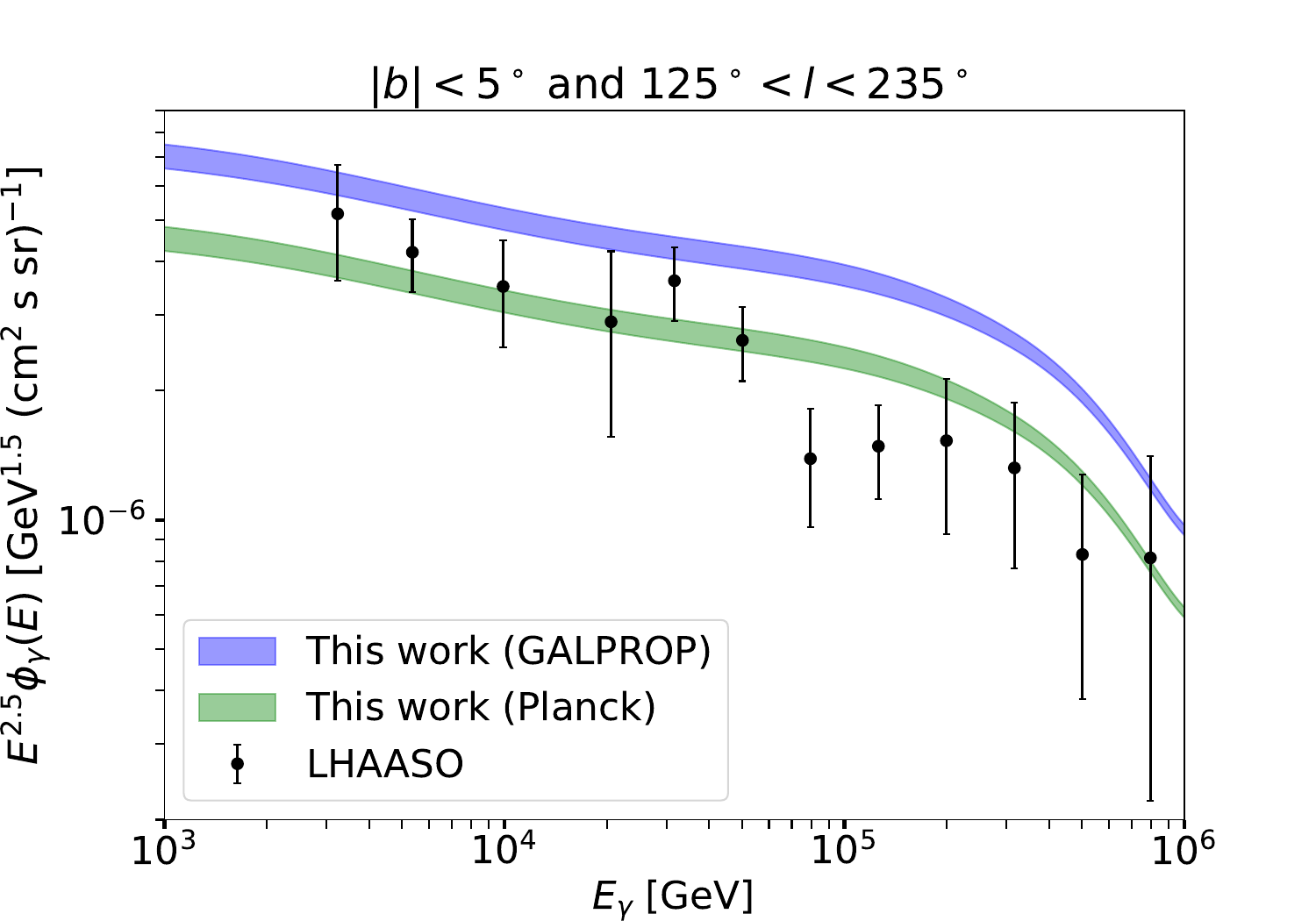}
\caption{Diffuse gamma-ray flux in LHAASO inner ($|b|<5^\circ$, $15^\circ<l<125^\circ$, top panel) and lateral ($|b|<5^\circ$, $125^\circ<l<235^\circ$, bottom panel) Galaxy regions. The $\gamma$-flux expectations obtained from our model are shown by blue (GALPROP gas model) and green (Planck gas model) shaded bands. Observational data by LHAASO \citep{Cao_2025a} are added with black points.}
\label{fig:lhaaso_flux_models}
\end{figure}

The diffuse $\gamma$-ray fluxes computed using the methodology discussed in the previous section are presented in Fig.~\ref{fig:lhaaso_flux_models} for both sky regions probed by LHAASO and after applying the same mask used by the collaboration in~\cite{Cao_2025a}. 
Two shaded bands, which represent our uncertainty in CR composition and contribution of heavy elements, illustrate our predictions based on different assumptions for the interstellar gas distributions. 
The blue band is obtained by using the gas distribution provided with the GALPROP code. 
The green band is based instead on the gas column density inferred from the 
Planck dust opacity ($\tau_D$) map.
\par The predictions obtained using the Planck dust map show a better agreement with LHAASO measurements, especially in the lateral region (\(|b| < 5^{\circ}\) and \(125^{\circ} < l < 235^{\circ}\)). 
In this sky region, the blue shaded band, derived from the GALPROP gas model, systematically exceeds the data points for gamma-ray energies \(E_\gamma > 30\) TeV. 
This trend is also observed in the inner region, with \(15^{\circ} < l < 125^{\circ}\); however, in this case, both predictions fail to adequately reproduce the LHAASO data below \(E_\gamma < 30\) TeV.
 
It is important to note that the observation of an excess of $\gamma$-rays, as recently discussed in \citet{Vecchiotti:2024kkz} and in the original LHAASO $\gamma$-ray papers \citep{Cao_2023,Cao_2025a}, may suggest the presence of a missing diffuse $\gamma$-ray component or a non-standard cosmic ray diffusion paradigm. However, the evidence of a deficit in observed emission represents a new and more puzzling feature.

We find this result cannot be attributed to the hadronic interaction model adopted for the diffuse flux calculations.
In order to show this, we compute the diffuse $\gamma$-flux expectation using other cross-section parametrizations in the literature.
Specifically, we consider Geant4, Pythia8, SIBYLL and QGSJET as defined in \citet{Kafexhiu_2014}) and we include the results in the Appendix. 
We see that the AAFrag cross-section gives the smaller (and softer) $\gamma$-ray predictions and, therefore, the tension with LHAASO $\gamma$-ray observations would increase for other parameterizations.

\begin{figure}
\centering
\includegraphics[scale=0.35]{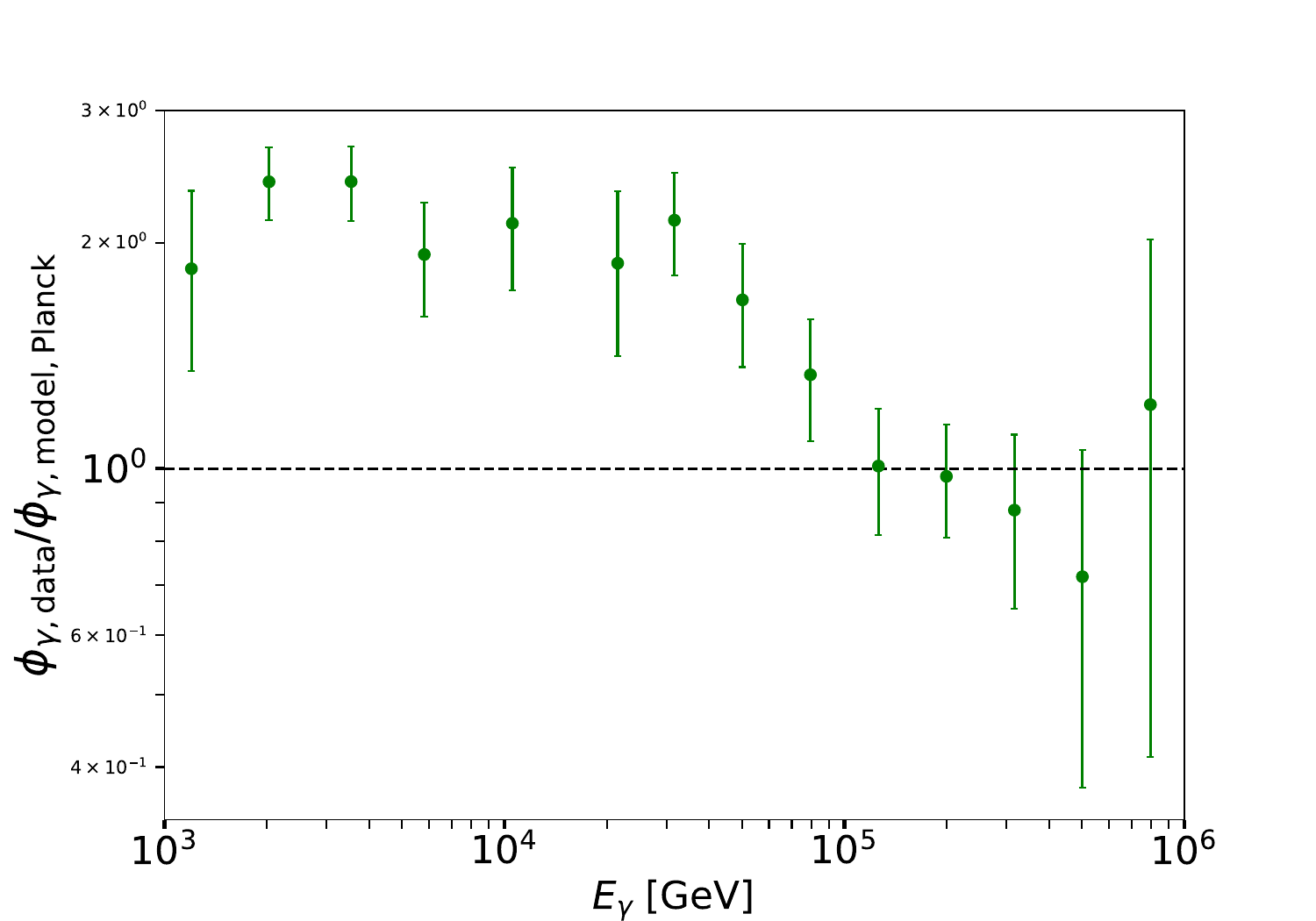}
\centering
\includegraphics[scale=0.35]{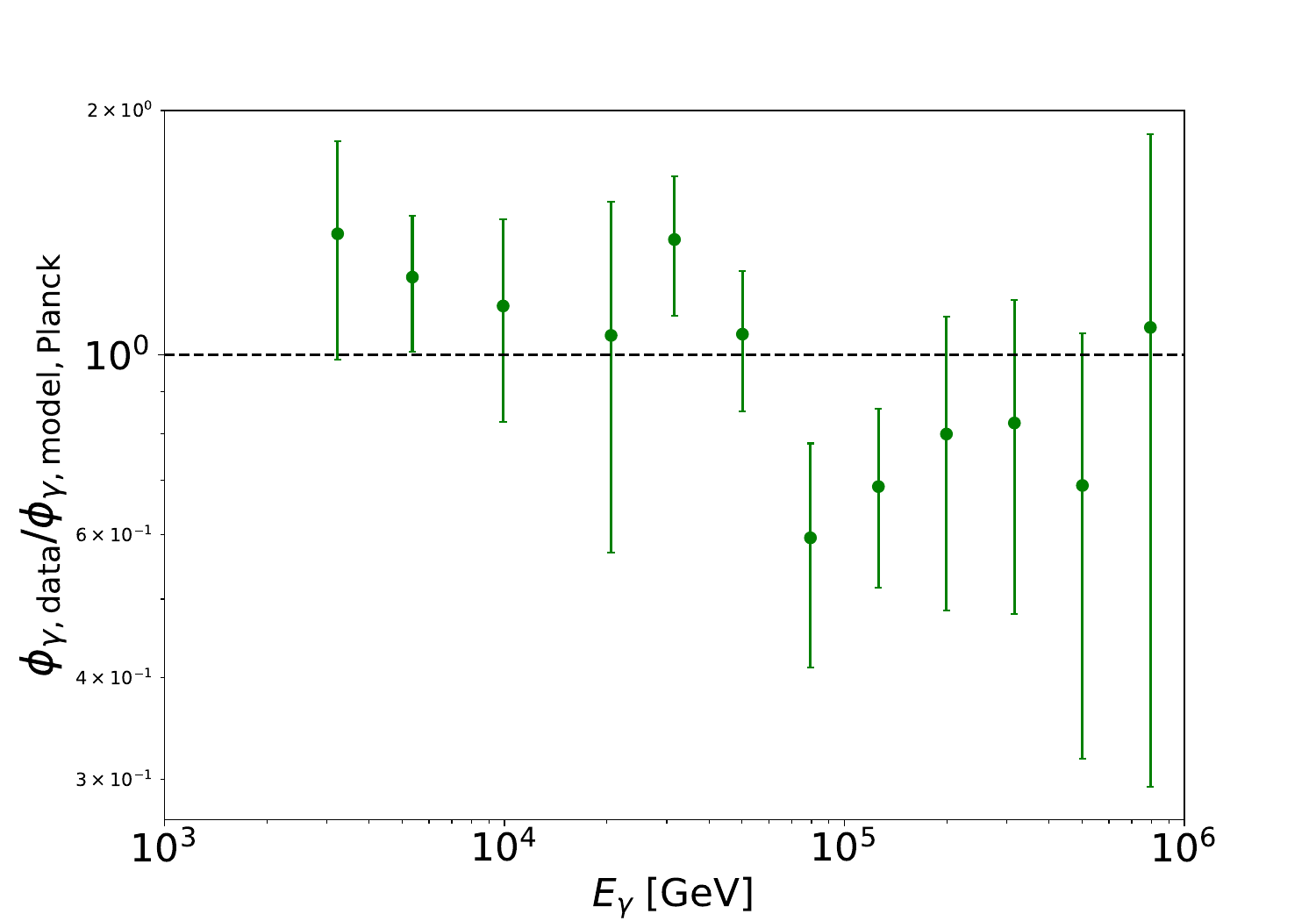}
\caption{Ratio of measured and expected $\gamma$-fluxes (for Planck gas distribution scenario) in LHAASO inner ($|b|<5^\circ$, $15^\circ<l<125^\circ$, top panel) and lateral ($|b|<5^\circ$, $125^\circ<l<235^\circ$, bottom panel) Galaxy regions as a function of gamma-ray energy. Dashed black line represents scenario were measurement data and expectations from our model perfectly agree.}
\label{fig:ratio_data}
\end{figure}

\par In order to better investigate the mismatch among data and predictions, we plot in Fig.~\ref{fig:ratio_data} their ratio as a function of the $\gamma$-ray energy in both LHAASO sky regions.
We take predictions obtained by using the Planck gas model as a reference, since they avoid overshooting LHAASO observational data in the lateral region.
We observe a clear deviation from the perfect agreement scenario (given by a constant ratio of $1$) in the inner Galaxy region and a less pronounced deviation in the lateral region. 
Notably, the discrepancy between data and predictions reflects not only in the overall normalization but also in the energy dependence of the ratios. 
%
%
Since diffuse $\gamma-$ray emission is essentially shaped by the CR proton spectrum (as explained in the previous section), this indicate a possible tension between $\gamma$-ray observations by LHAASO and the CR proton flux measured by LHAASO.

%
\begin{figure}
\centering
\includegraphics[scale=0.35]{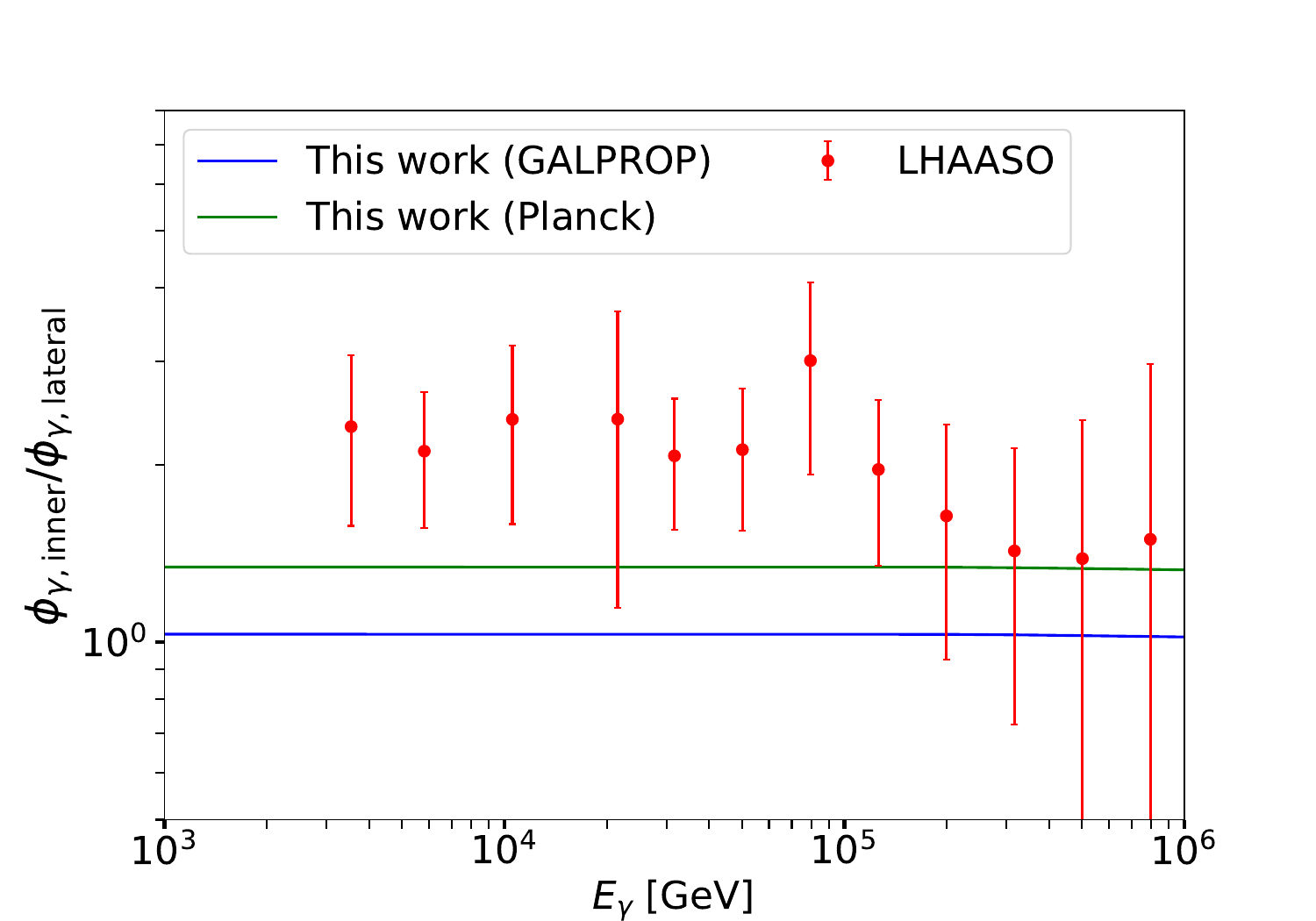}
\caption{Ratio of $\gamma$-fluxes in inner and lateral Galaxy regions as a function of gamma-ray energy. Ratio obtained from expectations by our model using GALPROP Galactic gas map \citep{Porter_2022} and Planck Galactic dust opacity map \citep{Planck2011} displayed with blue and green solid lines, respectively. Ratio computed with observation data by LHAASO \citep{Cao_2025a} shown with black scatter points.}
\label{fig:ratio_regions}
\end{figure}

Another important diagnostic that we consider is the ratio between the observational data of $\gamma$-fluxes in the inner and lateral regions, plotted in Fig.~\ref{fig:ratio_regions}. 
If we assume that the CR spectrum in the two regions is the same, 
this ratio is approximately energy independent, since we get
\begin{equation}
\frac{\phi_{\gamma,\, {\rm in}}(E_\gamma)}
{\phi_{\gamma,\, {\rm lat}}(E_\gamma)}
=
\frac{\int_{\rm in} d\Omega\; {\mathcal
  N}(E_\gamma,\hat{n}_{\gamma})} 
{\int_{\rm lat} d\Omega \; {\mathcal
  N}(E_\gamma,\hat{n}_{\gamma})}\,
\end{equation}
and the energy dependence of the function ${\mathcal N(E_\gamma,\hat{n}_{\gamma})}$ is mild in the standard scenario.
%
%
Thus, we expect that the ratios between observations performed in different sky regions can be fitted by a constant, apart for small deviations at very high energy due to the gamma-ray absorption. 
Moreover, in case of uniform CR density distribution, $g(r)=1$, the value of this constant is only determined by the amount of targets in two regions. 

In Fig.~\ref{fig:ratio_regions}, we show with a blue (green) line the expected ratio between the $\gamma$-ray fluxes from the two regions obtained by using GALPROP (Planck) gas distribution. 
In both cases, we see that predictions are lower than observational results, especially at energies below 100 TeV. 
The discrepancy with observational data appears, however, to be more significant when using the GALPROP gas model.


Finally, the observation of a energy-dependent flux ratio 
between the two sky regions could be a hint  for a spatial dependence of the CR spectral index in the Galaxy. 
Previous works \citep{Gaggero_2015, Acero_2016, Yang_2016, Pothast_2018, Pagliaroli:2016,Recchia_2016, Cerri_2017, Lipari_2018, Cataldo:2019qnz} have explored this possibility, based on the fact that Fermi-LAT $\gamma-$ray data at GeV energies suggest CR spectral hardening in the direction of the Galactic Center.
We note, however, that diffuse $\gamma$-ray observations by LHAASO do not distinguish between these spatial-dependent CR transport models and the conventional scenario (as was already noted in \citet{Vecchiotti:2024kkz}), being well compatible with a constant within uncertainties.

\section{Discussion}
\label{sec:disc}

We have computed the Galactic diffuse $\gamma$-ray flux expected by implementing the measurement of the proton flux by the LHAASO Collaboration. 
Namely, we have modeled the proton and helium fluxes, based on CR data from GeV to PeV energies and considered a conservative range for heavy nuclei (sub-dominant) contribution. 
We have considered different models of Galactic gas distribution and hadronic cross-section in the literature and found a persistent discrepancy between the predictions in all scenarios and the most recent LHAASO $\gamma$-ray flux observations. 
This disagreement is evident not only in the overall normalization but also in the spectral shape of the $\gamma$-flux. 


We have obtained an over-prediction of the $\gamma$-flux compared to data which is particularly evident in the lateral LHAASO observation region and when using GALPROP gas distribution.
We highlight that this tension is difficult to alleviate. 
An overall normalization difference can only be attributed to the uncertainty on the Galactic gas or variations of the CR density in the Galaxy. 
One possibility that has been discussed in \citet{Cataldo:2019qnz, Vecchiotti:2024kkz} is that the cosmic ray density is not uniform in the Galaxy but resembles the distribution of sources (mainly supernova remnants, SNR), motivated by a possible confinement in the proximity of the sources. 
However, including this variation in our analysis would cause a reduction of the gamma ray flux of up to 20$\%$ in the lateral region (in the most optimistic scenarios) while it would increase the prediction in the inner region, worsening the tension.

On the other hand, the discrepancy of the predicted spectral shape can be related to uncertainties on hadronic interaction models, $\gamma$-ray absorption, spatial dependence of the CR energy spectrum or the contribution of an additional diffuse component (e.g. a population of unresolved sources). 
Nevertheless, we have shown that the adoption of other cross-sections would increase the $\gamma$-flux while predicting a slightly harder slope. 
Similarly, an additional diffuse component contributing to the LHAASO measurement would require a lower $\gamma$-ray flux produced by CRs, especially in the inner region where a higher density of unresolved sources could be expected.
Therefore, both options would enlarge the discrepancy found in this work. 

Two alternatives that were not fully explored in this paper are the additional $\gamma$-ray absorption by interstellar radiation field (ISRF) and the possibility of a spatial-dependent CR energy spectrum. 
Absorption of $\gamma$-rays with energies $E_\gamma\gtrsim 30$ TeV, due to kinematical threshold for pair production, would require a population of background photons more energetic than CMB.
Their contribution is, however, believed to be small according to present estimates, see e.g.\citet{Vernetto:2016alq}. 
\par To assess the possibility of CR-spectrum spatial dependency, we have computed the ratio between the diffuse $\gamma$-ray emission measured by LHAASO in the two different sky regions. 
We note due to the large uncertainties, LHAASO data do not distinguish a hint of spatial dependent spectral indices in the TeV-PeV energy range.
The ratio among the current data set from the two regions, as reported in Fig.\ref{fig:ratio_regions}, is consistent with a constant value, which indicates that, in principle, the cosmic ray (CR) spectrum in both regions is the same. Nevertheless, the ratio predicted from both Galactic gas maps is considerably below the ratio obtained by observations at gamma-ray energies $E_\gamma \lesssim 10^2$ TeV. This difference in normalization could be attributed to a slightly higher density of gas or CRs in the inner region. However, this would increase the high-energy gamma-ray flux ($E_\gamma \gtrsim 10^2$ TeV) in the inner region, possibly leading to a predicted flux above the LHAASO observations.
\par Conversely, the ratios between the models and the data shown in Fig.\ref{fig:ratio_data} reveal a significant mismatch between the spectral shape of local CR and the measured gamma rays. This discrepancy is most noticeable in the inner region for gamma-ray energies $E_\gamma \lesssim 10^2$ TeV, while is the lowest in the lateral region.

One possible explanation is that the CR spectrum we are measuring may not be the one responsible for the observed gamma-ray emission. Specifically, the CR spectrum in the remainder of the Galaxy could be characterized by a different \textit{knee} location at around $E_p \sim 300$ TeV, which is compatible with the observed gamma break. This idea, partially explored in previous work \citep{Prevotat_2024}, could be directly tested in the future through gamma-ray observations.

Indeed we believe that a ratio between the diffuse gamma ray emission in different regions of the Galaxy reported directly by the LHAASO collaboration could offer more clear information by a significant reduction of the systematical uncertainties.

The uncertainty in the proton flux can potentially  affect our conclusions. Previous measurements by KASCADE~\citep{Antoni_2005, Finger_2011} reported a significantly lower proton flux than IceTop in the $\sim$PeV energy range. 
A recent work in Ref.~\citet{DeLaTorre_2025} highlighted that the $\gamma-$flux measured by LHAASO-KM2A was incompatible with the IceTop proton flux dataset while in good agreement with the KASCADE data.
Notably, a recent reanalysis of the KASCADE proton flux~\citep{Kuznetsov_2024} finds consistency with both IceTop and LHAASO measurements. If this updated dataset is confirmed by the KASCADE collaboration, then all charged cosmic-ray results would be mutually compatible, making the persistent tension with the diffuse gamma-ray observations even more puzzling.

Conversely, the helium flux lacks measurements between 100 TeV and a few PeV, providing another potential source of uncertainty in all models so far. 
Measurements of the cosmic ray composition (given by $\left< \ln A \right>$) by LHAASO together with their proton flux seem to suggest that the cosmic ray flux is dominated by protons between 1-10 PeV. 
These measurements do not exclude the possibility of a helium contribution at the knee lower than modeled in this work, potentially decreasing the diffuse $\gamma$-ray flux.
%
Dedicated helium flux measurements in the 100 TeV–PeV band are therefore essential~\citep{2024icrc.confE.413W}; once available, they will be incorporated into our framework in future work.

\section*{Acknowledgements}

The Authors thank Prof. Yuhui Li for providing the LHAASO mask needed to correctly perform the estimation of the diffuse emission. The work of GP and FLV is partially supported by  grant number 2022E2J4RK "PANTHEON: Perspectives in Astroparticle and
Neutrino THEory with Old and New messengers" under the program PRIN 2022 funded by the Italian Ministero dell’Universit\`a e della Ricerca (MUR) and by the European Union – Next Generation EU.

\section*{Data Availability}

All external data used in this work were obtained from the cited sources. Data produced in the analysis of this study are available upon reasonable request.



\bibliographystyle{mnras}
\bibliography{biblio} 





\bsp	
\label{lastpage}
\end{document}